\documentclass[conference]{IEEEtran}
\IEEEoverridecommandlockouts
\usepackage{cite}
\usepackage{amsmath,amssymb,amsfonts}
\usepackage{algorithmic}
\usepackage{graphicx}
\usepackage{textcomp}
\usepackage{xcolor}
\usepackage{booktabs}
\usepackage{multirow}
\usepackage[numbers]{natbib}
\usepackage{hyperref}
\usepackage{xfrac}

\def\BibTeX{{\rm B\kern-.05em{\sc i\kern-.025em b}\kern-.08em
    T\kern-.1667em\lower.7ex\hbox{E}\kern-.125emX}}

\makeatletter
\newcommand{\linebreakand}{
  \end{@IEEEauthorhalign}
  \hfill\mbox{}\par             
  \mbox{}\hfill\begin{@IEEEauthorhalign}
}
\makeatother
\begin{document}

\title{Pixel-Resolved Long-Context Learning for Turbulence at Exascale: Resolving Small-scale Eddies Toward the Viscous Limit\thanks{This manuscript has been authored by UT-Battelle LLC under contract DE-AC05-00OR22725 with the US Department of Energy (DOE). The US government retains and the publisher, by accepting the article for publication, acknowledges that the US government retains a nonexclusive, paid-up, irrevocable, worldwide license to publish or reproduce the published form of this manuscript, or allow others to do so, for US government purposes. DOE will provide public access to these results of federally sponsored research in accordance with the DOE Public Access Plan (https://www.energy.gov/doe-public-access-plan).}}
\iftrue
\author{\IEEEauthorblockN{Junqi Yin}
\IEEEauthorblockA{ 
\textit{Oak Ridge National Laboratory}\\
Oak Ridge, TN, USA \\
yinj@ornl.gov}
\and
\IEEEauthorblockN{Mijanur Palash}
\IEEEauthorblockA{ 
\textit{Oak Ridge National Laboratory}\\
Oak Ridge, TN, USA \\
palashmr@ornl.gov}
\and
\IEEEauthorblockN{M. Paul Laiu}
\IEEEauthorblockA{ 
\textit{Oak Ridge National Laboratory}\\
Oak Ridge, TN, USA \\
laiump@ornl.gov}
\linebreakand 
\IEEEauthorblockN{Muralikrishnan Gopalakrishnan Meena}
\IEEEauthorblockA{ 
\textit{Oak Ridge National Laboratory}\\
Oak Ridge, TN, USA \\
gopalakrishm@ornl.gov}
\and
\IEEEauthorblockN{John Gounley}
\IEEEauthorblockA{ 
\textit{Oak Ridge National Laboratory}\\
Oak Ridge, TN, USA \\
gounleyjp@ornl.gov}
\and
\IEEEauthorblockN{Stephen M.~de Bruyn Kops}
\IEEEauthorblockA{ 
\textit{University of Massachusetts Amherst}\\
Amherst, MA, USA \\
debk@umass.edu}
\linebreakand 
\IEEEauthorblockN{Feiyi Wang}
\IEEEauthorblockA{ 
\textit{Oak Ridge National Laboratory}\\
Oak Ridge, TN, USA \\
fwang2@ornl.gov}
\and
\IEEEauthorblockN{Ramanan Sankaran}
\IEEEauthorblockA{ 
\textit{Oak Ridge National Laboratory}\\
Oak Ridge, TN, USA \\
sankaranr@ornl.gov}
\and
\IEEEauthorblockN{Pei Zhang}
\IEEEauthorblockA{ 
\textit{Oak Ridge National Laboratory}\\
Oak Ridge, TN, USA \\
zhangp1@ornl.gov}
}
\fi
\maketitle

\begin{abstract} 
Turbulence plays a crucial role in multiphysics applications, including aerodynamics, fusion, and combustion. Accurately capturing turbulence's multiscale characteristics is essential for reliable predictions of multiphysics interactions, but remains a grand challenge even for exascale supercomputers and advanced deep learning models. The extreme-resolution data required to represent turbulence, ranging from billions to trillions of grid points, pose prohibitive computational costs for models based on architectures like vision transformers. To address this challenge, we introduce a multiscale hierarchical Turbulence Transformer that reduces sequence length from billions to a few millions and a novel RingX sequence parallelism approach that enables scalable long-context learning. We perform scaling and science runs on the Frontier supercomputer. Our approach demonstrates excellent performance up to 1.1 EFLOPS on 32,768 AMD GPUs, with a scaling efficiency of 94\%. To our knowledge, this is the first AI model for turbulence that can capture small-scale eddies down to the dissipative range. 
\end{abstract}

\section{Problem Overview}
\label{sec:ProblemOverview}



Turbulent flows are central to a wide range of applications, such as aerodynamics, energy generation, fusion energy sciences, and environmental sciences. 
In these applications, turbulence may be intentionally introduced to enhance efficiency or naturally occurring, adding complexity to scientific understanding. 
A better understanding of turbulence is crucial for the simulation and design of next-generation energy devices and advancing fundamental science knowledge.

Fluid turbulence remains a grand challenge due to its inherently multiscale and stochastic nature \cite{Pope_2000}. 
In multiphysics applications, its complexity is further compounded by non-linear interactions with other physical processes like chemical reactions \cite{bilger2005paradigms} and electromagnetism \cite{snyder2001electromagnetic}.
Capturing all the relevant scales accurately requires first-principle-based simulations with extremely high resolutions (e.g., 35 trillion grid points \cite{yeung2025gpu}), which can take days to perform even on advanced exascale computers.
Low-fidelity physics-based modeling is often used in practical applications, but requires turbulence closure models that, today, cannot be reliably applied to new flow configurations. 
In contrast, once trained, data-driven deep learning (DL) models can significantly accelerate the time-to-solution. In particular, emerging foundation model (FM) approaches hold promise for providing a single, generalizable solution across diverse turbulence tasks, e.g., providing physical insights, closure modeling, and fast surrogate modeling.  

Training such a turbulence FM inevitably faces multiscale challenge arising from the extremely high-resolution data. 
Representing all the multiscale features in the data requires an expressive model architecture and prohibitive computing resources. 
Existing scientific FM work has made great progress toward AI models at impressive sizes, such as ClimaX \cite{climax} with 115 million parameters, Aurora  \cite{aurora} with 1.3 billion parameters, and ORBIT \cite{orbit} with 113 billion parameters. However, all these transformer-based models overlook the orthogonal direction regarding the underlying `AI model resolution' with a multiple-pixel patch size (e.g., $8\times8$ for ClimaX and $4\times4$ for Aurora) and hence miss the small-scale features.
On the other hand, existing single-purpose DL for turbulent flows struggle with similar challenges: they are limited to down-sampled data, e.g., two-dimensional (2D) slices and small three-dimensional (3D) blocks, and to cost-effective model design options like large patch sizes in model training. These limitations eventually lead to DL models that match simple statistics well, while missing key small-scale physics that are crucial to model correctness and accuracy. 
To capture those small-scale features, we need a pixel-resolved transformer model, which is computationally prohibitive with existing algorithms and computing resources.

In this work, we tackle the challenge with two novel techniques: a physics-inspired multiscale hierarchical Turbulence Transformer and the highly efficient RingX sequence parallelism algorithm. We demonstrate the effectiveness of our approach on the DOE Oak Ridge Leadership Computing Facility's (OLCF) Frontier supercomputer, achieving both cutting-edge scaling performance and scientific advancement in two high-fidelity turbulence problems:
\paragraph{Forced Homogeneous Isotropic Turbulence (HIT)}
HIT is an idealized turbulence model with statistically equivalent properties in all directions and spatially uniform. It is a well-accepted benchmark in direct numerical simulation (DNS) studies, often used to push the computing limits and advance our understanding of turbulence phenomena, such as scaling laws and the energy cascade between eddies at different scales \cite{yeung2025gpu}.
Forced HIT data from the Johns Hopkins Turbulence Database (JHTDB)~\cite{li2008public} is a widely used open-source test case in the community. The dataset originates from a DNS with a time sequence using a spatial resolution of $1024^3$ grid points, where the energy was injected at large scales by keeping total energy at the wave number $\kappa\leq2$ constant. Energy then cascades from large to small scales within the inertial range, following the classic $-5/3$ scaling law ($\sim\kappa^{-5/3}$), and is dissipated by viscosity within the dissipative range. While many previous studies using DL models have successfully resolved the scales within the inertial range, none have captured the eddies approaching the Kolmogorov scale in the dissipative range in 3D high-resolution turbulence data. 
\paragraph{Stably Stratified Turbulence (SST)}
SST is a model used to understand turbulent flows influenced by a background density stratification, resulting in highly intermittent and anisotropic behavior across multiple scales. Studying SST has important implications for multiple fields (e.g., pollution mitigation and deep sea mining) and is also valuable for advancing fundamental turbulence theory (e.g., turbulent/non-turbulent interfaces, intermittency, and anisotropic multiscale energetics introduced by buoyancy). In classical turbulence theory, the Reynolds number $\mathit{Re}$ provides a single control parameter for characterizing the flow. However, when gravitational forces become important, the resulting SST depends strongly on multiple additional parameters including the Prandtl number $\mathit{Pr}$ (ratio of the fluid's kinematic viscosity to density diffusivity) and the Froude number $\mathit{Fr}$ (ratio of buoyancy to inertial time scales). Simulating fully-resolved flows across this multi-dimensional parameter space, and modeling the resulting dynamics, is thus of critical importance for better understanding turbulence in a variety of geophysical and industrial settings, but requires exascale computing resources to achieve the necessary resolution to capture the vast range of dynamically-relevant length scales.

\begin{figure*}[t]
\centering 
\includegraphics[width=.9\linewidth]{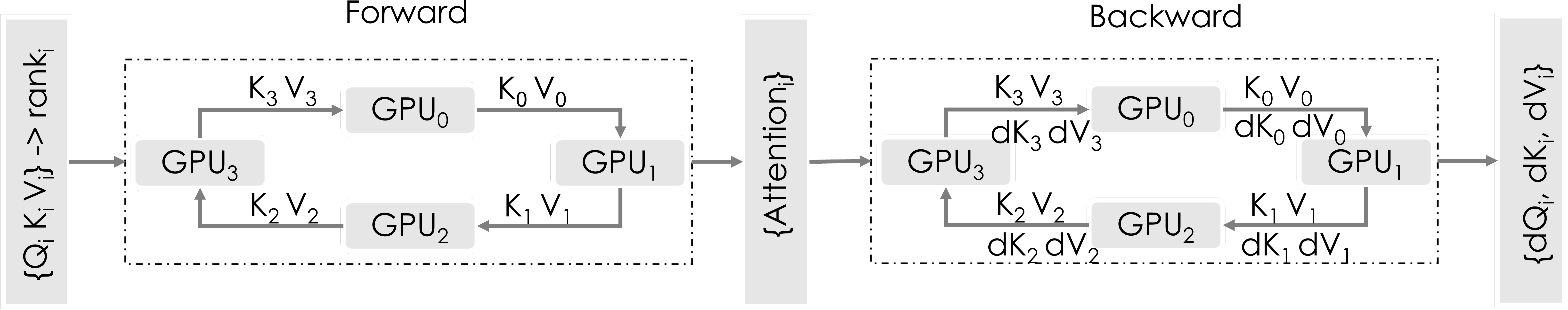}
\caption{Illustration of Ring attention. Each GPU calculates its local attention with respect to a query $Q_i$ by exchanging Key and Value tensors, ($K_{i-1}, V_{i-1}$) and ($K_{i+1}, V_{i+1}$), with neighbors in a communication ring.}
\label{fig:ring}
\end{figure*}
\noindent \textbf{Long-context challenge} 
AI modeling at the pixel-level resolution is required to tackle the multiscale physics in turbulence system, which indicates the need for billions to trillions context length learning for ViT. The self-attention mechanism, which is the key factor behind the transformer architecture's outstanding performance, is defined as 
\begin{align}
    A^s =  \sum^S_{i=1}\text{Softmax}(Q_s\cdot K_i^\intercal) V_i\,. \label{eq:attn}
\end{align}
where $\{(Q_s,K_s,V_s),s=1,\cdots S\}$ are the query, key, and value vectors, respectively, which have the size equal to the hidden dimension and are evenly split across the attention heads, and $S$ represents the context length.
The calculations of the attention score (i.e., $Q_s\cdot K_i^\intercal$) and the subsequent projection over $V_i$ are both quadratic in computational and memory complexity. This makes long-context training a major challenge.

Many efforts have been devoted to parallelize the attention calculation, including the sequence parallelism in DeepSpeed-Ulysses \cite{ulysses} and the context parallelism in Megatron \cite{megatron}. The terms `sequence parallelism' and `context parallelism' are used interchangeably in this context. However, both methods have limited scalability due to their parallelization strategies---restricted by the limited number of attention heads to partition and the memory bottleneck caused by the \texttt{Allgather}'ed sequence, respectively. Ring Attention \cite{ring} is the first parallel method that supports unlimited sequence length via block-wise computation, and it has been widely adopted for extreme long context training \cite{llama3, gradient}. However, Ring Attention relies on information exchange among neighboring GPUs, which is not the most efficient way of utilizing a HPC platform with high-speed interconnect \cite{allgather}; in such environments, \texttt{Allgather} is often faster than simple ring-based point-to-point communication.

\section{State of the art}

\noindent \textbf{SOTA in scientific FMs and DL for turbulence} 
Typical FMs for physical systems are limited to small, 2D systems \cite{mccabe2023multiple, zhang2024matey}. 
Those extended to large-scale, mostly for weather and climate applications \cite{climax, aurora, orbit}, focus on scaling up the model size, rather than exploiting model's capabilities in capturing data features via long-context attention. 
No FM currently exists that targets high-fidelity turbulence data. 
However, task-specific DL have shown promise in various turbulence applications, including super-resolution reconstruction \cite{fukami2019super}, 
turbulence generation \cite{lienen2023zero}, 
closure modeling \cite{li2025transformer}, and spatiotemporal prediction \cite{wang2020towards}. 
The dominant model architecture has shifted from convolutional neural networks to transformers for the latter's improved expressibility and scalability. 

However, the latest transformer-based studies that use high-fidelity turbulence data still involve significant data downsampling: for example, going from 3D to 2D for populating samples due to either limited data \cite{fukami2024single} or computational constraints \cite{liu2025multi}. 
Recent work \cite{sardar2024concerning} highlights the drawbacks of such downsampling, particularly its failure to capture small-scale structures in 2D turbulence, underscoring the need to process full-resolution, long-sequence turbulence data.
 Computer vision approaches such as hierarchical ViTs \cite{liu2022swin} reduce sequence lengths but without accounting for the scales in physical systems. 
 In contrast, simulation-inspired approaches such as adaptive tokenization \cite{zhang2024matey} also reduce sequence length by leveraging spatial inhomogeneity—but not to the extent required for high-resolution turbulence data.

\noindent \textbf{SOTA in long context}
The current state of the art method for parallel attention is Ring Attention. As illustrated in Fig.~\ref{fig:ring}, by dividing the attention computation into blocks---each device calculates ``local attention" of $Q$, $K$, $V$ chunks---Ring Attention \cite{ring} is the first parallel attention method that supports unlimited context length. The devices form a logical ring, where $K$ and $V$ are exchanged between neighbors via point-to-point communications. The local results are integrated, while each $K, V$ pair travels through the entire ring. The ring communication topology has been shown to be effective \cite{horovod} in bandwidth-limited systems, and the Ring Attention is the state-of-the-art method for ultra long context (e.g., millions of tokens). However, on HPC platforms equipped with premium interconnects, such as most AI data centers designed for large language model (LLM) training, collective communications are more efficient by leveraging the network hierarchy (e.g., the tree topology). There are efforts \cite{DISTFLASHATTN, tree-attn} focusing on LLM-specific optimizations, but they either depend on using a specific framework (e.g., DeepSpeed) \cite{DISTFLASHATTN} or work solely for inference \cite{tree-attn}. 

\section{Innovation Realized}
We present two novel methods, RingX parallel attention and the multiscale hierarchical Turbulence Transformer architecture, necessary to achieve pixel-level-resolution context length for resolving multiscale physics in turbulence systems.

\subsection{RingX Parallel Attention}
\begin{figure*}[t]
\centering 
\includegraphics[width=.9\linewidth]{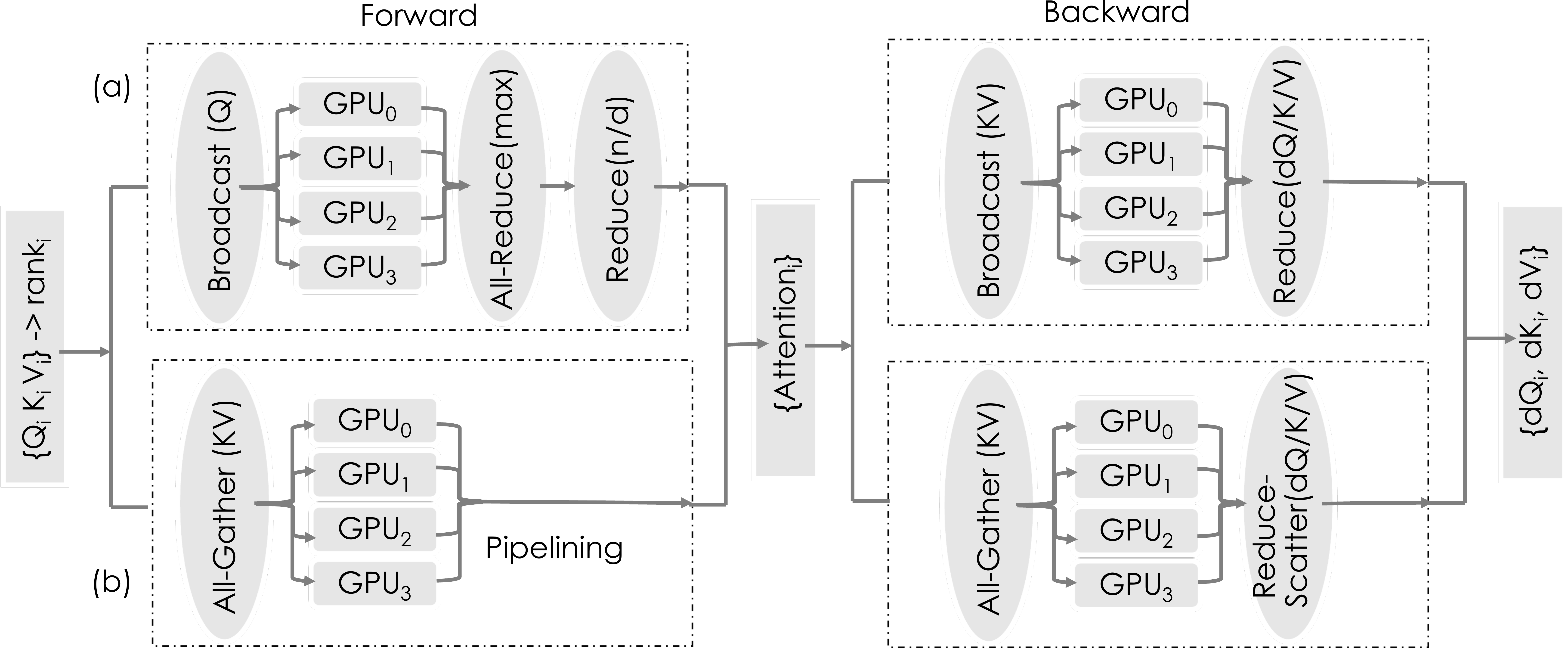}
\caption{Illustration of RingX Attention. Each GPU calculates its local attention with respect to a query $Q_i$ by (a) broadcasting the query and calculating the local attention with each $(K, V)$ simultaneously or (b) gathering the key and value tensors ($K, V$) and adding up attention locally.}
\label{fig:ringx}
\end{figure*}

Here, we propose RingX, a parallel attention method based on Flash Attention \cite{flash}, and achieve better performance and scalability than Ring Attention, with improved workload partitioning, a more efficient communication scheme, and better overlap of communication and computation, as illustrated in Fig.~\ref{fig:ringx}.

\noindent \textbf{Workload partition} In Ring Attention, the computation of Eq.~\eqref{eq:attn} is divided into $\{K, V\}$ blocks. For a query $Q_n$, the attention $A^n$ is then the combination of those local attention results.  
The local chunk of $Q_s$ on each GPU will be used to calculate the local output with every chunk of $\{K, V\}$, which requires the point-to-point exchange of $K$ and $V$ among neighboring GPUs in this logic ring (Fig.~\ref{fig:ring}). In practice, the \texttt{Softmax} is calculated using the so-called log-sum-exp (LSE) trick \cite{flash} for numerical stability, and the attention formula is re-written as, 
\begin{align}\textstyle
   A^s &= \frac{\sum^S_{i=1}\text{exp}{\left(\text{LSE}_s - \text{Max}(\text{LSE}_s)\right)} \cdot V_i}{\sum^S_{s=1}\text{exp}{\left(\text{LSE}_s- \text{Max}(\text{LSE}_s)\right)}}  \label{eq:tree}\\
    \text{LSE}_s &= \textstyle \log\sum_j \text{exp}{(e^s_{j})} \\
    e^s_j &= \textstyle Q_s \cdot K_j^\intercal \label{eq:score}
\end{align}
where $e^s_j$ is the attention score between token $j$ and $s$ in the sequence. Observing that the summations in the numerator and denominator of Eq.~\eqref{eq:tree} are also associative, an alternative parallel scheme is to partition the workload with $N$ GPUs (each GPU dealing with a chunk of $\frac{S}{N}$) by  
\begin{align}
    num &= \textstyle
    \sum_i^N \left(\mathrm{loc\_out}_i \text{exp}(\mathrm{loc\_lse}_i - \mathrm{lse_{max}})\right)  \notag \\ 
    den &= \textstyle
    \sum_i^N \text{exp}(\mathrm{loc\_lse}_i - \mathrm{lse_{max}})  \label{eq:ringx1}
\end{align}
where $\mathrm{loc\_out}_i$ and $\mathrm{loc\_lse}_i$ are the local attention output and LSE on the sequence chunk $S_i$, and can be calculated using Flash Attention \cite{flash}.  

\noindent \textbf{Communication scheme} Instead of iterating over the $\{K, V\}$ chunks in a communication ring, every GPU can work on the same query simultaneously. The communication pattern changes to a \texttt{Reduce} collective to sum up the local contributions to $num$ and $den$. There is an extra \texttt{Allreduce} operation to obtain the global maximum on LSE, i.e., $\mathrm{lse}_{max}$, which is needed for the numerical stability of the exponential calculation. Note that the above scheme is for a single query. In practice, the parallel attention method is applied with context parallelism, i.e., the $Q$, $K$, and $V$ tensors are chunked in the sequence dimension. In that case, the calculation of Eq.~\eqref{eq:ringx1} needs to be performed for each chunk in total $N$ steps, and the $\frac{num}{den}$ gives the corresponding attention. Specifically, in the forward pass, each $Q$ chunk is broadcasted to all GPU ranks and the local attention outputs are calculated simultaneously; in the backward pass, $K$, $V$ chunks are broadcasted in order to compute their gradients simultaneously and then reduced over $N$ ranks, while the gradient for $Q$ is accumulated for each rank over $N$ steps. We denote this method as Ring$X_a$, as illustrated in Fig.~\ref{fig:ringx}.

Considering the \texttt{Broadcast} is typically optimized \cite{allgather} towards small message sizes (e.g., control messages), another collective scheme is to use \texttt{Allgather} to first collect all $K, V$ chunks, and then each GPU adds up each contribution to the attention locally. Correspondingly, in the backward pass, the gradients need to be obtained via \texttt{ReduceScatter}. We denote this method as Ring$X_b$. Note that this approach is different from the context parallel in Megatron-core \cite{megatron}, where the $Q, K, V$ tensors are all-gathered before a single sequential Flash Attention calculation.

\noindent \textbf{Communication and computation overlap} We further optimize the algorithm by overlapping the communication with computation via pipelining the adjacent two steps with asynchronous collectives. Specifically, we initiate the \texttt{Broadcast} for the next iteration's Q at the end of the current step, while the model continues local computations. The synchronization is only performed when the local results are strictly needed. The memory usage is slightly increased because two buffers are used in a ping-pong manner to avoid overwriting data that is still in use.   

\noindent \textbf{Memory and communication cost analysis} 
Regarding the communication requirement, RingX is either comparable to or less expensive than Ring. The straightforward \texttt{Allgather} of Q, K, V tensors in Megatron results into a message volume per GPU of $6\frac{S}{N}hb$, while the \texttt{Send-Recv} in Ring attention reduces the message volume by $\frac{1}{3}$ because only K and V are exchanged. Ring$X_a$ further reduces the payload by about $75\%$ (the largest communication is for Q only), while the \texttt{Allgather} in Ring$X_b$ has the same communication volume as in Ring attention, but using HPC-oriented collectives.         

In terms of the memory requirement, both Ring and RingX benefit from the optimized Flash Attention implementation \cite{flash}, which reduces the memory complexity from quadratic to linear with respect to sequence length. Depending on the communication patterns, the memory required per GPU can remain constant when the total number of GPUs for the context parallelism increases proportionally with the sequence length. This enables unlimited context for Ring$X_a$, although the computation cost still grows quadratically with the sequence length. Due to the usage of \texttt{Allgather}, Megatron and Ring$X_b$ are limited to the sequence length that a single GPU can fit, while Ring$X_b$ still has a smaller memory footprint compared to Megatron. The detailed requirements are listed in Table~\ref{tab:perf}. Our methods and analyses are platform-agnostic and generally applicable to attention-based applications.      

\begin{table}[]
\centering
\caption{Communication and memory requirements for RingX and other parallel attention methods}\label{tab:perf} 
\begin{tabular}{lllll}
\toprule
\multirow{2}{*}{Method} & \multicolumn{2}{l}{Communication (forward)}                                                                                                       & \multirow{2}{*}{Memory} & \multirow{2}{*}{Context} \\
                      & Protocol                                                                             & Volume                                                     &                         &                          \\ \midrule
Megatron              & all-gather(qkv)                                                                      & 6$\frac{S}{N} h b$                                                     & O($Shb$)                  & limited                  \\
Ring           & send-recv(kv)                                                                        & 4$\frac{S}{N} h b$                                                   & O($\frac{S}{N} h b$)              & unlimited                \\ \hline
Ring$X_a$             & \begin{tabular}[c]{@{}l@{}}broadcast(q)\\ all-reduce(lse)\\ reduce(num,den)\end{tabular} & \begin{tabular}[c]{@{}l@{}}$\frac{S}{N} h b$\\ + 4$\frac{S}{N}b$ \end{tabular} & O($\frac{S}{N} h b$)              & unlimited                \\ \hline
Ring$X_b$                & all-gather(kv)                                                                       & 4$\frac{S}{N} h b$                                                   & O($Shb$)                  & limited                  \\ \bottomrule
b: batch size & h: hidden size & \multicolumn{2}{c} S: sequence length \\
\multicolumn{3}{l} N: number of GPUs for context parallelism 
\end{tabular}
\end{table}

\subsection{Turbulence Transformer}
\begin{figure}[t]
\centering 
\includegraphics[width=1.\linewidth]{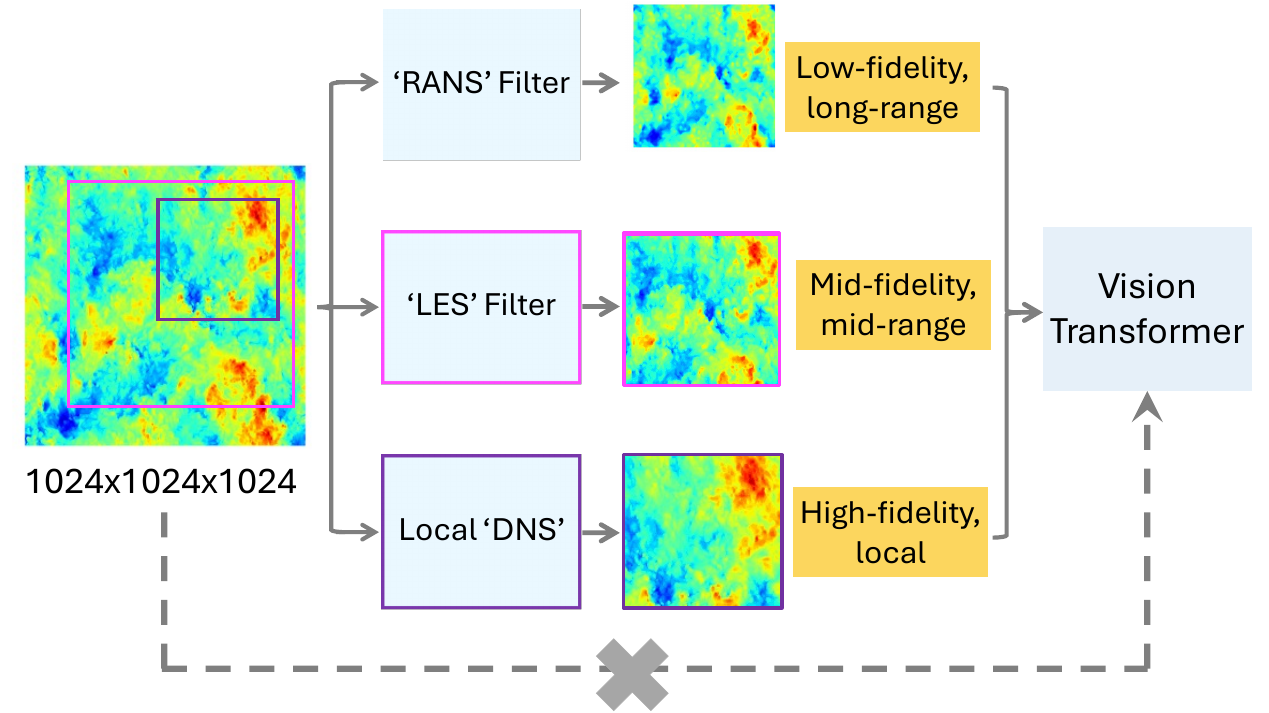}
\caption{Illustrative diagram of physics-guided multi-fidelity Turbulence Transformer,
which enables efficient ingestion of high-resolution 3D turbulence data while preserving accuracy.
}
\label{fig:diagram-tt}
\end{figure}

Sequence parallelism alone is insufficient to capture all scales of turbulence, as it would require attention over sequences with billions to trillions of tokens---well beyond the current computational capabilities, even for exascale computers.
To address this, we introduce a multiscale hierarchical Turbulence Transformer,  inspired by principles from physics-based modeling (illustrated in Fig.~\ref{fig:diagram-tt}). Conventional computational fluid dynamics (CFD) of turbulent flows consider three regimes: 1) DNS, resolving eddies at all scales, is the most accurate but the most expensive; 2) Reynolds Averaged Navier-Stokes (RANS) simulations, solving only mean fields and modeling all unresolved turbulent motions, offer the fastest solutions at the lowest fidelity; and 3) between the two, large eddy simulations (LES) resolve the large eddies while modeling the smaller ones.
Analogously, the Turbulence Transformer partitions attention over the extremely long input sequence into smaller, scale-specific sequences: high-fidelity for the local, fine-scale interactions; low-fidelity for the long-range, large-scale global physics; and the middle-fidelity for interactions at intermediate scales.

Let $\bold Y\in\mathbb{R}^{D\times H\times W\times 4}$ denote the state of the turbulence system at a grid resolution $D\times H\times W$ with four variables representing the three velocity components ($V_x, V_y, V_z$) and pressure $P$.
For the spatiotemporal learning task, a ViT model is trained to predict a future state from a sequence of $M$ input states. More precisely, we aim to learn a function $\mathbf{f}_{\mathbf{w}}$ such that 
\begin{equation}
\bold Y^{t+t_{\text{lead}}} \approx \bold f_\bold w(\bold Y^{t-M+1},\ldots,\bold Y^{t}; t_{\text{lead}}),    
\end{equation}
 where $t_{\text{lead}}$ denotes the prediction lead time.
In the standard ViT with a patch size of $[p_z, p_x, p_y]$, the input token/patch sequence length is
    $L=M\times D/p_z\times H/p_x\times W/p_z$.
To resolve the smallest features, we require $p_x=p_y=p_z=1$. 
For DNS cases such as JHTDB (in which $D=W=H=1024$), this resolution results in a sequence length on the order of billions---a scale that is computationally prohibitive.

In the Turbulence Transformer, we leverage the insights from physics-based modeling to decompose this computationally prohibitive task into multi-fidelity representations, each with a manageable sequence length. Fig.~\ref{fig:diagram-tt} illustrates an architecture with the three fidelity modes: 
\begin{itemize}
    \item The low-fidelity mode targets global long-range physics with small-scale features filtered out with a Gaussian kernel $\mathcal{F}_1(\cdot)$ of size $k_1^3$ via convolution, i.e., $\mathcal{F}_1(\bold Y)\in\mathbb{R}^{\sfrac{D}{k_1}\times \sfrac{H}{k_1}\times \sfrac{W}{k_1}\times 4}$.
    \item The mid-fidelity mode is composed of a Gaussian filter $\mathcal{F}_2(\cdot)$ of kernel size $k_2^3$ and a subdomain selection operator $\mathcal{C}_2(\cdot)$ of size $D_2\times H_2\times W_2$ and centered at $[x_{c2},y_{c2},z_{c2}]$, to focus on the mid-range physics, i.e., $\mathcal{F}_2(\mathcal{C}_2(\bold Y))\in\mathbb{R}^{\sfrac{D_2}{k_2}\times \sfrac{H_2}{k_2}\times \sfrac{W_2}{k_2}\times4}$. 
    \item The high-fidelity mode represents local physics with $\mathcal{C}_3(\cdot)$ of size $D_3\times H_3\times W_3$, i.e., $\mathcal{C}_3(\bold Y)\in\mathbb{R}^{D_3\times H_3\times W_3\times 4}$.
\end{itemize}
The multi-fidelity structure is characterized by three hyperparameters: filter size $[k_1,k_2, k_3]$, the location $[x_{ci},y_{ci},z_{ci}]$, and the size of each fidelity mode $[D_i, H_i, W_i]$ ($i=1,2,3$). The values of these parameters are determined by the physics and computational constraints. We implement the Turbulence Transformer in a model called Matey and evaluate the performance. In our tests, the locations are uniformly randomly sampled, and we will report the filter size and the fidelity model size ratio $[D/D_1, D/D_2, D/D_3]$. We note that $D/D_i=H/H_i=W/W_i$ in our test cases.

\section{How Performance Was Measured}

\noindent \textbf{Hardware platform} Our experiments are performed on the Frontier supercomputer. Each Frontier node is equipped with four AMD Instinct MI250X GPUs with dual Graphics Compute Dies (GCDs) and one third-generation EPYC CPU. A GCD is viewed as an effective GPU, and we use GCD and GPU interchangeably in the following discussion. All four MI250Xs (eight effective GPUs) on a node are connected using 100 GB/s Infinity Fabric (200 GB/s between 2 GCDs of MI250X). Nodes are connected via a Slingshot-11 interconnect with 100 GB/s of bandwidth. Frontier consists of 9408 nodes for a total of 75,264 effective GPUs, each equipped with 64GB of high-bandwidth memory (HBM2E).

\noindent \textbf{Software environment} Our software stack is based on PyTorch v2.4.0 built against ROCm v6.1.3 on Frontier. Our parallel attention implementation will be made publicly available.

\subsection{FLOPs counts}
\begin{table}[]
\centering
\caption{Computations in attention kernel.}\label{tab:flops} 
\begin{tabular}{lcc}
 \textbf{Module} & \textbf{GEMM Size} & \textbf{FLOPs}  \\ 
 \toprule
 Input to Q/K/V  & $(b, S, h)\times(h, h)$  & $6bSh^2$ \\
 Attention Score & $({ba}, S, \frac{h}{a}) \times ({ba}, \frac{h}{a}, S)$ & $2bS^2h$ \\
 Attn over Value & $({ba}, S, S) \times ({ba}, S, \frac{h}{a})$ & $2bS^2h$  \\
 Linear Projection & $(bS, {h}) \times ({h}, h)$ & $2bSh^2$ \\
\bottomrule
b: batch size & h: hidden size & S: sequence length \\
a: number of heads
\end{tabular}
\end{table}

\begin{figure}[t]
\centering 
\includegraphics[width=.95\linewidth]{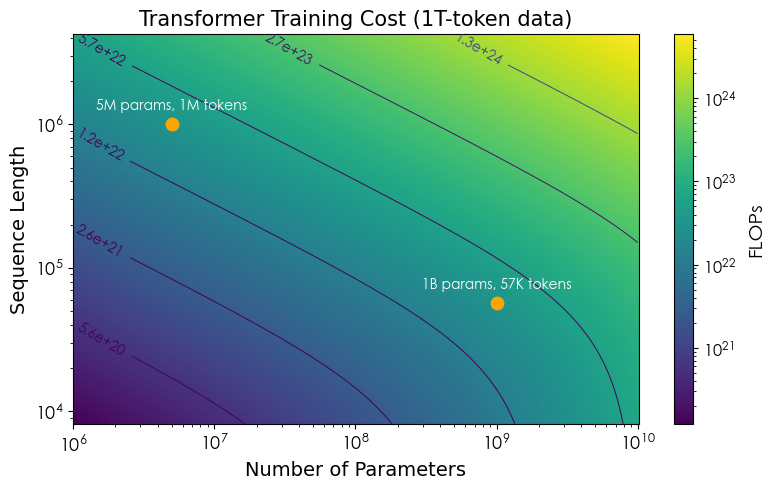}
\caption{The training cost estimate (for 1T tokens) as a function of the model parameter and the sequence length.}
\label{fig:cost}
\end{figure}

Our RingX methods calculate the exact attention as defined by Eq.~\eqref{eq:attn} (within numerical rounding for low precision arithmetic), and hence the required floating point operations (FLOPs) are the same. For both ViT and GPT applications, a typical attention block includes input transformation (i.e., multiplication of input tensor with Q, K, V tensors, respectively), attention score calculation (Eq.~\eqref{eq:score}), attention value calculation (Eq.~\eqref{eq:tree}), and linear projection.     
Given the sequence length $S$, batch size $b$, and hidden size $h$, the FLOPs for the forward-pass attention matrix (See Table~\ref{tab:flops}) computation follows 
\begin{align}
    F = 4bSh^2 (S/h+2), \label{eq:flops}
\end{align}
where $b\cdot S$ is the total input tokens and $h^2$ (multiplying the number of layers) represents the model parameters. Therefore, when $S$ is comparable to $h$, the total model training FLOPs is proportional to data and model sizes regardless of input sequence length. In the case of long context, i.e., $S\gg h$, the computation is dominated by the attention calculation and the total budget grows linearly with the sequence length for the fixed model and data sizes. For example, training a $10\times$ longer context length requires the same amount of extra computation as that of a $10\times$ larger model, given the same data. As shown in Fig.~\ref{fig:cost}, for a fixed data size, the training cost for a 5M parameter model with 1M token sequence length is about the same as for a 1B parameter model with 57K squence length. Hence the long-context training is expensive, and the optimization of parallel attention is important and can significantly reduce the cost.

\subsection{System telemetry} In addition to computational metrics such as run time and FLOP counts, we track the energy and GPU utilization via system telemetry. The \texttt{rocm-smi} reports the average energy usage at the default one  millisecond sampling interval. By mapping the cyclic power, memory, and utilization data with the training iterations, we estimate the per iteration statistics. Considering each training step follows the same computational workload, this provides a good estimator, based on which the total energy consumption and hence computational efficiency (in GFLOPS/Watt) can be calculated.     

\subsection{Application metrics}

We consider two turbulence applications: HIT and stratified Taylor-Green cases.
In HIT simulations \cite{jhtdb}, energy is injected into large-scale eddies at low wave numbers, cascades to small-scale eddies at high wave numbers, and dissipates in the dissipative range dominated by viscosity, following the classic Kolmogorov's 1941 scaling theory. 
The key turbulence descriptors are the energy spectrum, the energy dissipation rate, enstrophy, and the skewness and kurtosis of velocity gradients. The energy spectrum $E(\kappa)$ shows energy distribution at different wave numbers $\kappa$, while the dissipation rate $\varepsilon=2\nu S_{ij}S_{ij}$ and enstrophy $\Omega=\omega_i\omega_i/2$ describe the structure of small-scale turbulence \cite{yeung2012dissipation}.
\footnote{Here, $\omega_i$ is the vorticity, $\nu$ is the viscosity, and $S_{ij}$ is the strain rate tensor.} Enstrophy quantifies the intensity of vortical motions, while dissipation describes local straining motions related to energy loss. The skewness (the third moment) represents the asymmetry level, while the kurtosis (the fourth moment) measures the tailedness of a distribution. 
A deeper understanding of these concepts is essential for both advancing fundamental turbulence theory and modeling multiphysics interactions, particularly in applications such as turbulent combustion \cite{chen2011petascale}.
However, no DL model has demonstrated such capabilities yet in high $\mathit{Re}$ 3D turbulence.
We will evaluate model accuracy with respect to these three quantities. 

The stratified Taylor--Green vortex \cite{riley03} is crucial for understanding the transition to turbulence, with stratified layers developing, shearing against each other, erupting into turbulence, and then dissipating and decaying.  
It is an extremely difficult problem to model because of the continuously evolving, spatially varying, patches of growing and decaying turbulence. We will evaluate model accuracy regarding the time history of kinetic energy and potential energy in the evolving process.

\section{Performance Results}
We present the scaling performance, followed by the scientific results.  
\begin{table}[]
\centering
\caption{Matey model architecture.}\label{tab:model}
\resizebox{0.5\textwidth}{!}{
\begin{tabular}{lllllll}
\toprule
Model  & Emb  & Patch     & TT with 3 modes     & time steps & SeqLen & \#Params \\
  &   &      & (filter; mode size ratio)     & &  &  \\\midrule
Small  & 192  & (2, 2, 2) & (1, 4, 8); (8$\times$, 2$\times$, 1$\times$)  & 4       & 1M     & 5M       \\
Medium & 384  & (1, 1, 1) & (1, 4, 8); (8$\times$, 2$\times$, 1$\times$)  & 2       & 4M     & 21M      \\
Large  & 960  & (1, 1, 1) & (1, 4, 8); (8$\times$, 2$\times$, 1$\times$)  & 1       & 2M     & 176M     \\
XL     & 3072 & (1, 1, 1) & (2, 8, 16); (8$\times$, 2$\times$, 1$\times$) & 4       & 1M     & 1.3B     \\ \bottomrule
TT: Turbulence Transformer\\
\end{tabular}
}
\end{table}
\subsection{Scaling}
Our scaling experiments are performed with several model sizes (ranging from Small with 5M parameters to XL with 1.3B parameters) and sequence lengths (1M, 2M, and 4M tokens). The specific model architectures are listed in Table~\ref{tab:model}. 
The computational cost for training a ViT model at sequence lengths of millions of tokens are substantial, as shown in Fig.~\ref{fig:cost}. For the same amount of data, training Matey-Small (5M parameter model with 1M sequence length) is more expensive than training the 1B parameter ORBIT model (sequence length of 2K), which is among the previous largest scale ViTs.  
We will compare our RingX approach with Ring Attention, and demonstrate the end-to-end application performance in the following.   

As indicated by Eq.~\eqref{eq:flops}, for training with long sequences, the required computation (FLOPs) is dominated by the attention kernel, the scaling of which is the key to the application scalability. 

\begin{figure}[t]
\centering 
\includegraphics[width=1.\linewidth]{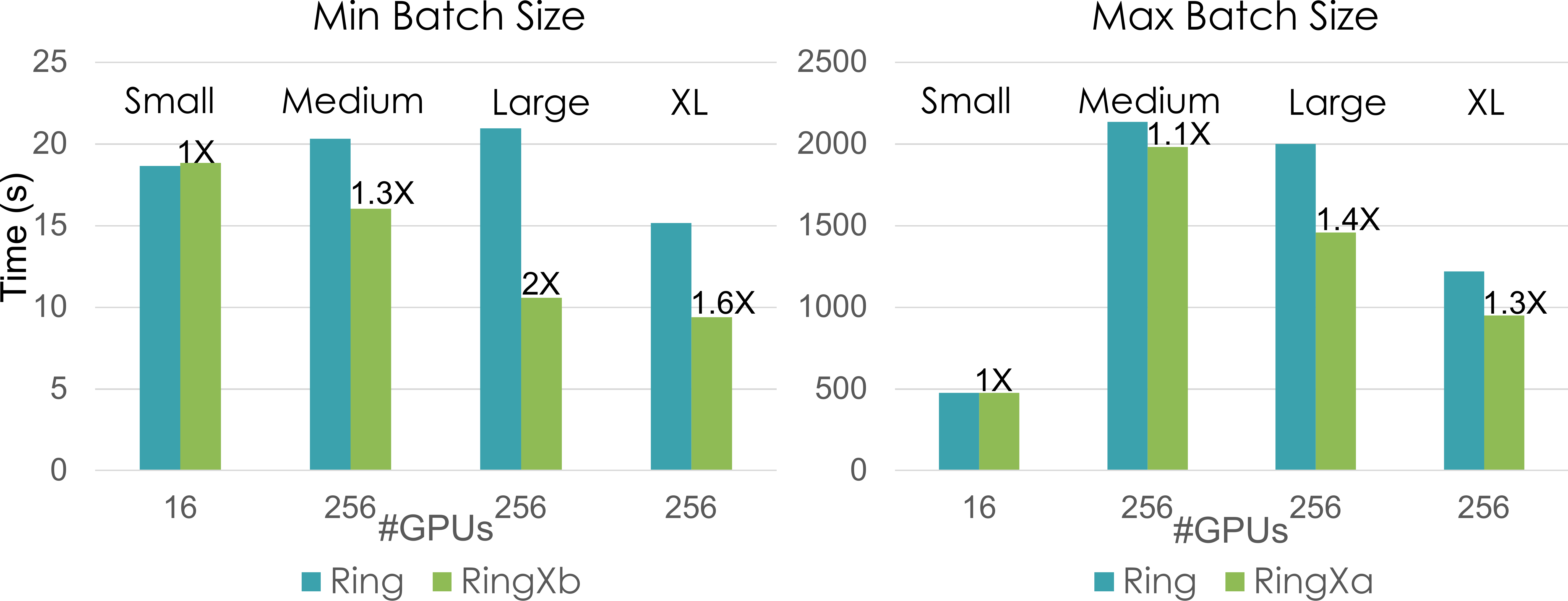}
\caption{The attention kernel performance (runtime) for Matey architectures. The context parallel size for Matey-Small is 16 and for the rest is 256, with smallest (Min) to largest (Max) batch sizes. The speedup for RingX over Ring is marked in each case.}
\label{fig:kernel-scaling1}
\end{figure}

\noindent \textbf{Kernel scaling} Ring Attention relies on the overlapped communication and computation to hide the communication overheads, but, in practical settings, its point-to-point communication often incurs the more significant cost. As shown in Fig.~\ref{fig:kernel-scaling1}, compared to Ring attention, RingX performs better for moderate to large model sizes, and achieves up to a $2\times$ speedup. For the 5M-parameter model (which is used as a baseline and often too small for ViT), when only 16 GPUs are used to parallelize the 1M sequence (i.e., 64K for each GPU), RingX and Ring perform similarly. As the scale of the context parallelism increases (for the purpose of either fitting larger models into the memory and/or reducing the runtime), RingX significantly outperforms Ring.         

\begin{figure}[t]
\centering 
\includegraphics[width=1.\linewidth]{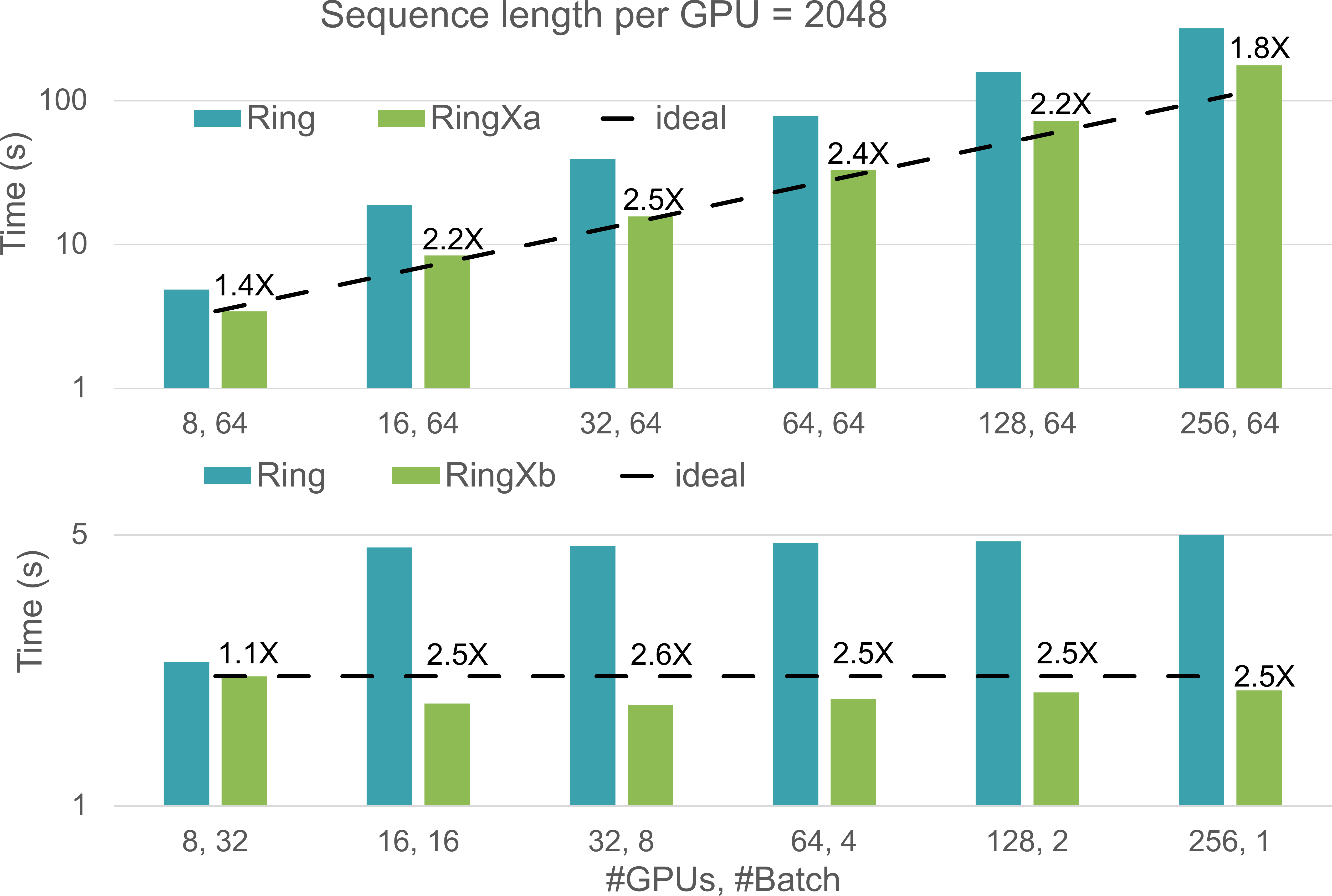}
\caption{The kernel scaling of RingX versus Ring attention. }
\label{fig:kernel-scaling2}
\end{figure}

In Fig.~\ref{fig:kernel-scaling2}, we show the runtime scaling of RingX, compared to Ring. The sequence length per GPU is fixed at 2K. For Ring$X_a$ (better for large batch sizes), the per GPU batch size is also fixed (64) while the number of GPUs increases from 8 to 256. All GPUs are in a single context parallel group, i.e., the total sequence length for the attention calculation is the multiplication of the number of GPUs with 2K. Because compute load scales quadratically (Eq.~\eqref{eq:flops}) with the sequence length, the ideal scaling of the runtime then grows linearly with the number of GPUs. Across all scales considered, RingX outperforms Ring by up to $2.5\times$.   

For Ring$X_b$ (better for small batch sizes), the total batch size is kept constant, i.e., the per GPU batch size reduces correspondingly with the increase of the number of GPUs (also the total sequence length). The ideal scaling curve is a flatten line. A similar speedup of $2.5\times$ for RingX over Ring is obtained at scale. The slight super-linear scaling is likely due to the performance boost of \texttt{Allgather} in Ring$X_b$ for larger message sizes.      

\begin{figure}[t]
\centering 
\includegraphics[width=.95\linewidth]{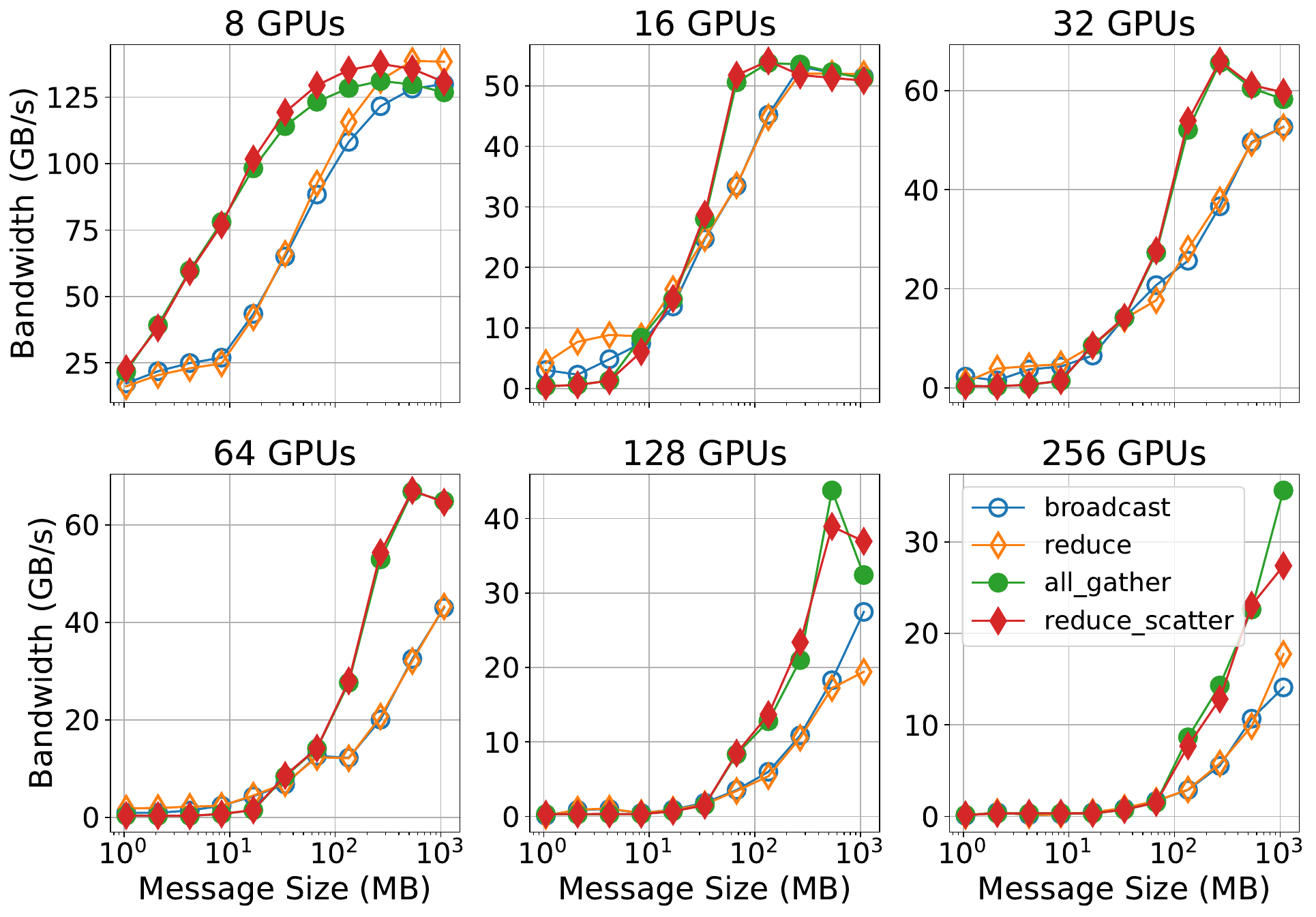}
\caption{The communication bandwidth (GB/s) for \texttt{Broadcast}, \texttt{Reduce}, \texttt{Allgather}, \texttt{ReduceScatter} using 8 to 256 GPUs on Frontier.}
\label{fig:rccl}
\end{figure}

To further understand the performance behavior of RingX, we measure the communication bandwidth of the underlying communication library (RCCL) on Frontier. As shown in Fig.~\ref{fig:rccl}, for small messages (e.g., small embedding size such as Matey-Small), the bandwidth for the collectives used in RingX is relatively low. As the message size (i.e., the embedding and model size) increases, the advantage of RingX grows with the bandwidth of the collectives. Additionally, \texttt{Allgather} and \texttt{ReduceScatter} employed in Ring$X_b$ performs better than \texttt{Broadcast} and \texttt{Reduce} used in RingX for the message size larger than hundreds of MB and the scale larger than 16 GPUs.      

\begin{figure}[t]
\centering 
\includegraphics[width=1.\linewidth]{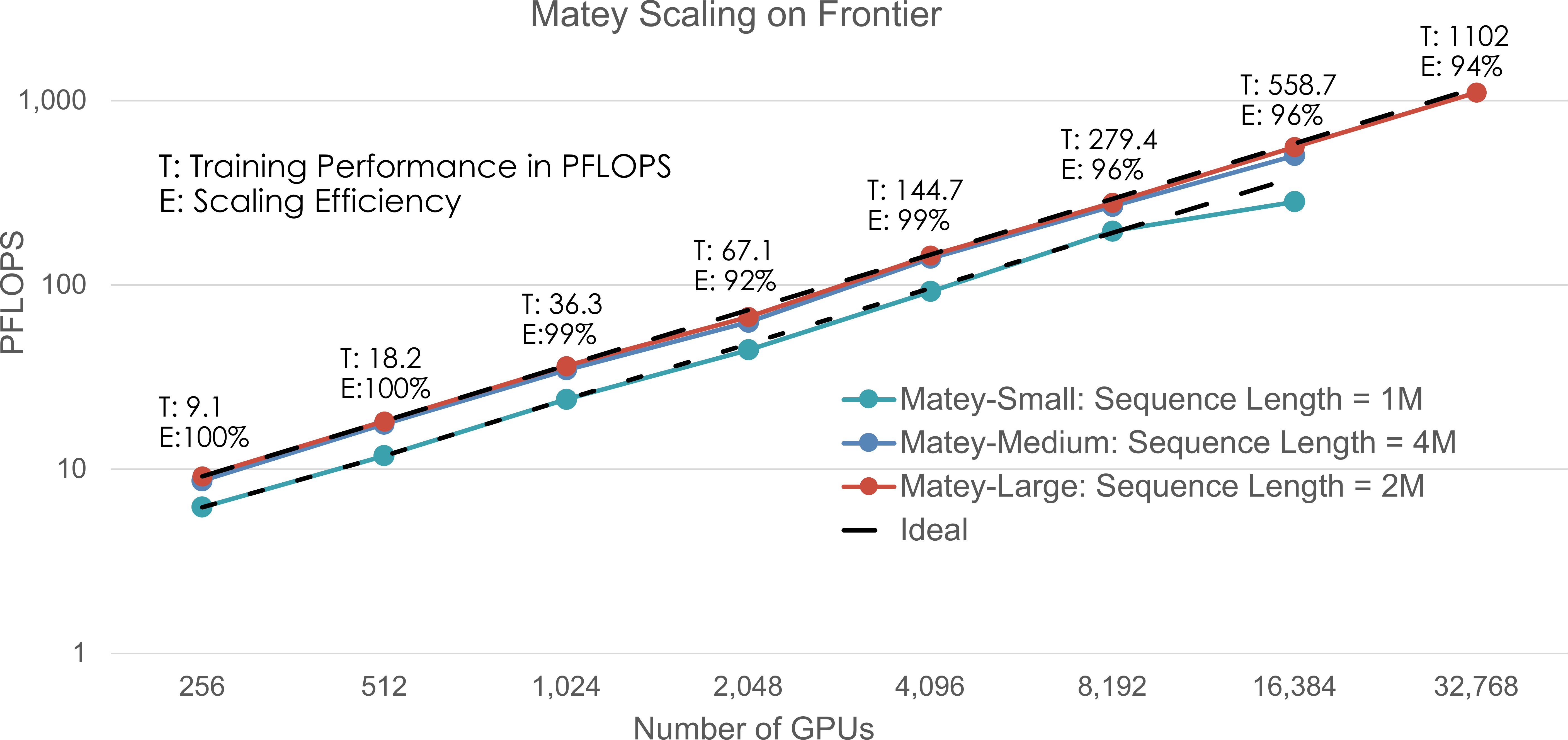}
\caption{The scaling performance (PFLOPS) of training Matey models on the Frontier Supercomputer.} 
\label{fig:app-scaling}
\end{figure}

\noindent \textbf{Application scaling} 
With the parallel attention kernel optimized, we evaluate Matey training at full system scale. In Fig.~\ref{fig:app-scaling}, we show the training throughput in PFLOPS from 256 to 32,768 GPUs for Matey-Small, Matey-Medium, and Matey-Large, respectively. The measurements are based on the FLOP counts from Eq.~\eqref{eq:flops} and the measured runtime. This is a conservative estimate of performance as it includes only attention-related FLOPs, though the actual total would be only marginally higher.
We report the performance numbers starting from 256 GPUs because that is the minimum scale required to accomodate in 4M-token and 2M-token sequence lengths for Matey-Medium and Matey-Large, respectively. RingX is used for the parallel attention, where the local attention on each GPU is calculated with Flash Attention using mixed float32-bfloat16 precision. The specific run configurations are listed in Table~\ref{tab:config}. 

Since the computational budget in FLOPs scales linearly with the number of parameters and quadratically with sequence length, Matey-Large---with about $8\times$ more parameters and $2\times$ shorter sequence length than Matey-Medium---is the most expensive. At 256 GPUs, the training throughputs are 9.1, 8.6, and 6.2 PFLOPS for Matey-Large, Matey-Medium, and Matey-Small, respectively. All three models scale nearly linearly to 8,192 GPUs. At 16,384 GPUs, the scaling efficiency of Matey-Small drops to 75\%, , likely due to its higher communication-to-computation ratio, making it the most sensitive to communication overhead among the three. Matey-Large achieves a sustained 1.1 EFLOPS training throughput using 32,768 GPUs, with a scaling efficiency of 94\%. In comparison, the ORBIT 115M model (sequence length 1024) was reported at 613 PFLOPS at the same scale.           

\begin{figure}[t]
\centering 
\includegraphics[width=0.9\linewidth]{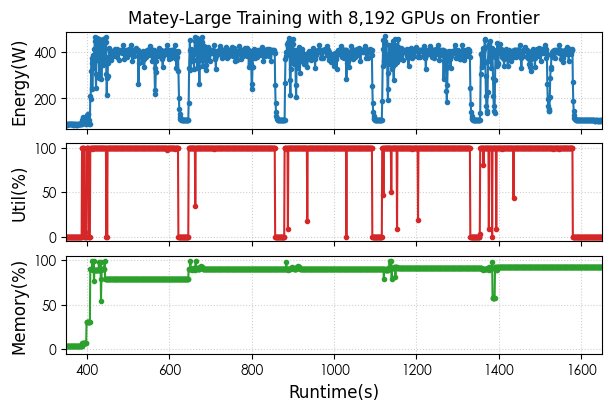}
\caption{The trace of GPU power, memory, and utilization during the first five training steps.  }
\label{fig:power}
\end{figure}

The telemetry data for the GPU power, memory, and utilization were also collected during the training. In Fig.~\ref{fig:power}, the trace (of only one GPU for clarity) for the first five batch steps of training Matey-Large using 8,192 GPUs are plotted. Because of the size of input data (multiple time steps of images 
 and each with $1000^3$ pixels) and relatively complex pre-processing, there are small gaps in between batch steps for the power and utilization curves. The average power consumption during the training is about 358 Watt, and hence the computational efficiency for training Matey-Large is 188 GFLOPS/Watt. In comparisons, the Frontier power efficiency for double precision application is reported \cite{green500} around 62 GFLOPS/Watt, which indicates our mixed-precision Matey training is about $3\times$ more efficient. The performance numbers for all three models are listed in Table~\ref{tab:config}.

\begin{table}[]
\centering
\caption{Matey parallel configurations and scaling performance. CP and DP are for context and data parallelism, respectively.}\label{tab:config}
\resizebox{0.5\textwidth}{!}{
\begin{tabular}{ccccccc}
\toprule
\multirow{2}{*}{Model} & \multirow{2}{*}{\begin{tabular}[c]{@{}c@{}}\# CP\\ Size\end{tabular}} & \multirow{2}{*}{\begin{tabular}[c]{@{}c@{}}\# DP\\ Size\end{tabular}} & \multirow{2}{*}{\begin{tabular}[c]{@{}c@{}}Batch\\ Size\end{tabular}} & \multirow{2}{*}{PFLOPS} & \multicolumn{2}{c}{Efficiency} \\
                       &                                                                        &                                                                        &                                                                       &                         & GFLOPS/Watt      & Scaling     \\ \midrule
Small                  & 16                                                                     & 1024                                                                   & 1024                                                                  & 282                     & 124              & 74\%        \\
Medium                 & 256                                                                    & 128                                                                    & 128                                                                   & 503                     & 177              & 91\%        \\
Large                  & 256                                                                    & 128                                                                    & 128                                                                   & 1101                    & 188              & 94\%        \\ \bottomrule
\end{tabular}}
\end{table}

\subsection{Forced HIT}

\begin{figure}[t]
\centering 
\includegraphics[width=0.9\linewidth]{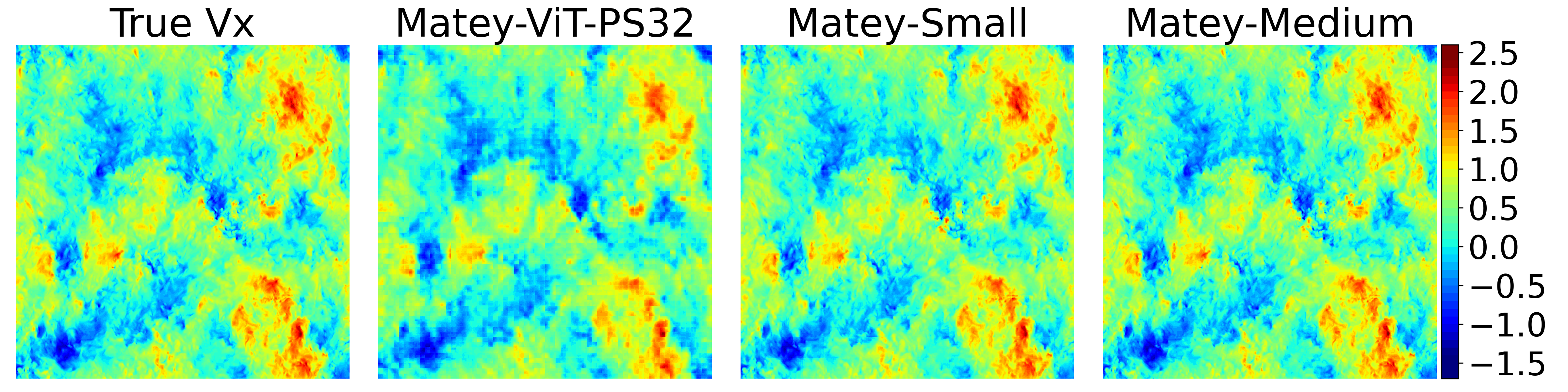}
\caption{Contours of velocity component $V_x$ at a cross-section from DNS and Matey predictions under three configurations.}
\label{fig:ucontours}
\end{figure}

In this experiment, we train Matey to predict next-step solution fields with state variables $\textbf{Y}^{n+1}=[V_x, V_y, V_z, P]$. 
The model takes as input the solutions from the previous $M=4$ steps. 
We evaluate three model configurations in Matey: the Matey-Small and Matey-Medium models from Table~\ref{tab:model}, together with a ViT at a patch size of $32^3$ with the same embedding size as Matey-Small (Matey-ViT-PS32). For Matey-Medium, we increase the input steps from 2 to 4 and adjust the Turbulence Transformer configuration to: filter size of $[1, 8, 16]$ and mode size ratio of $[16\times, 2\times, 1\times]$ to keep the same sequence length as Matey-Small.
 We obtained 100 time sequences with resolution $1024^3$ from JHTDB \cite{jhtdb} with the Taylor-scale $\mathit{Re}_\lambda\approx433$ and constructed 96 samples, using 80\% for training and the remaining 20\% for testing. 
Fig.~\ref{fig:ucontours} compares the contours of $V_x$ of a test sample with predictions from three model configurations. Clearly, the two Matey configurations with Turbulence Transformer that allow for smaller patch sizes capture much finer scales. 

\begin{figure}[t]
\centering 
\includegraphics[width=0.6\linewidth]{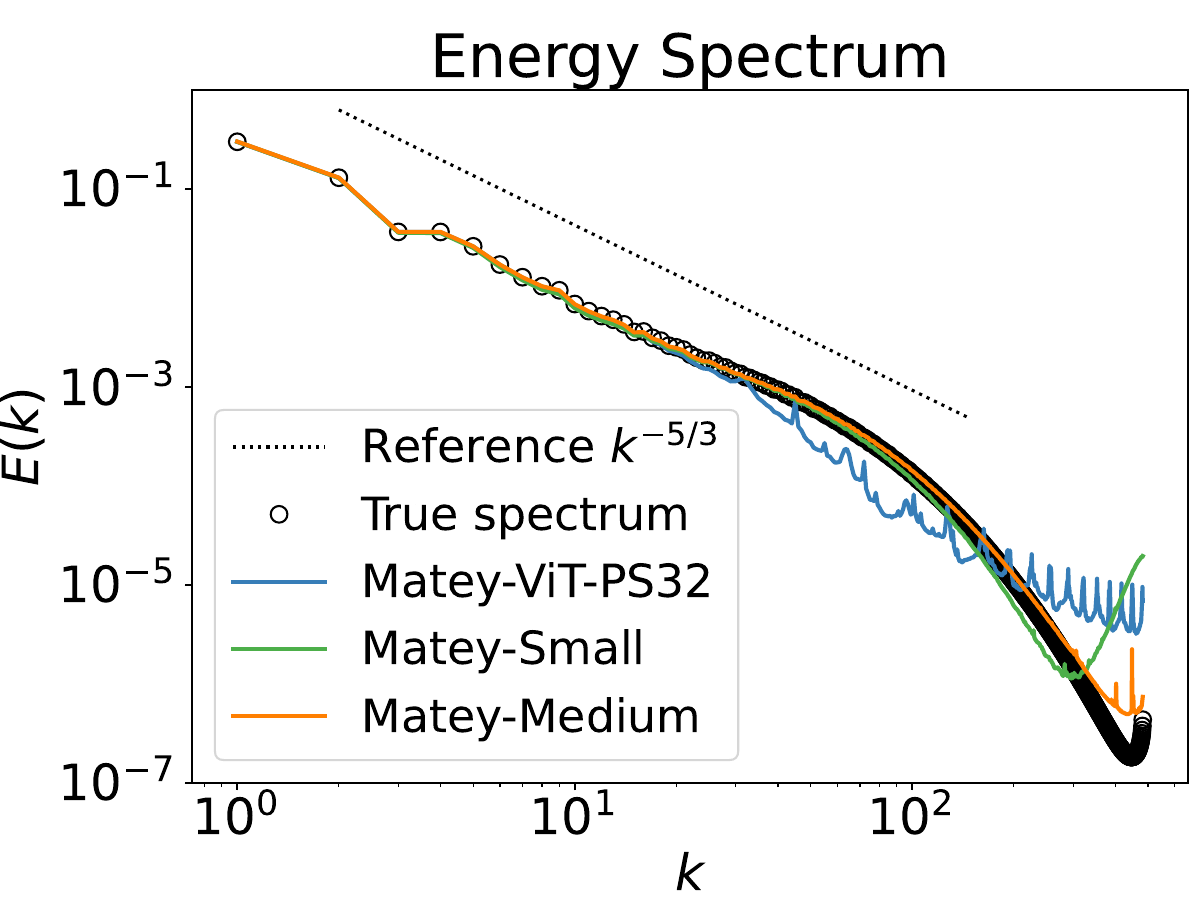}
\caption{Turbulence Energy Spectrum showing Matey capturing small-scale turbulence down to the dissipative range.}
\label{fig:espec}
\end{figure}

\noindent \textbf{Energy Spectrum} Fig.~\ref{fig:espec} compares the kinetic energy spectra derived from the predicted velocity fields of the three models, Matey-ViT-PS32, Matey-Small, and Matey-Medium, with the true spectrum from DNS data.
The dotted line indicates Kolmogorov's $-5/3$ scaling, with $E(\kappa)\sim \kappa^{-5/3}$, in the inertial range where kinetic energy is transferred from large eddies to small eddies. All three Matey configurations accurately capture the energy distribution in this regime.    
Beyond the inertial range lies the dissipative range, where kinetic energy is dissipated at small scales.
Models with large patch sizes fail to capture the energy distribution in this high-wavenumber regime. In contrast, Matey-Medium, with a patch size of $1^3$, successfully captures the full spectral range down toward the viscous limit.
To our knowledge, this represents the first demonstration of a DL model capable of resolving all relevant scales in 3D homogeneous isotropic turbulence at large $Re$.

\begin{figure}[t]
\centering 
\includegraphics[width=0.8\linewidth]{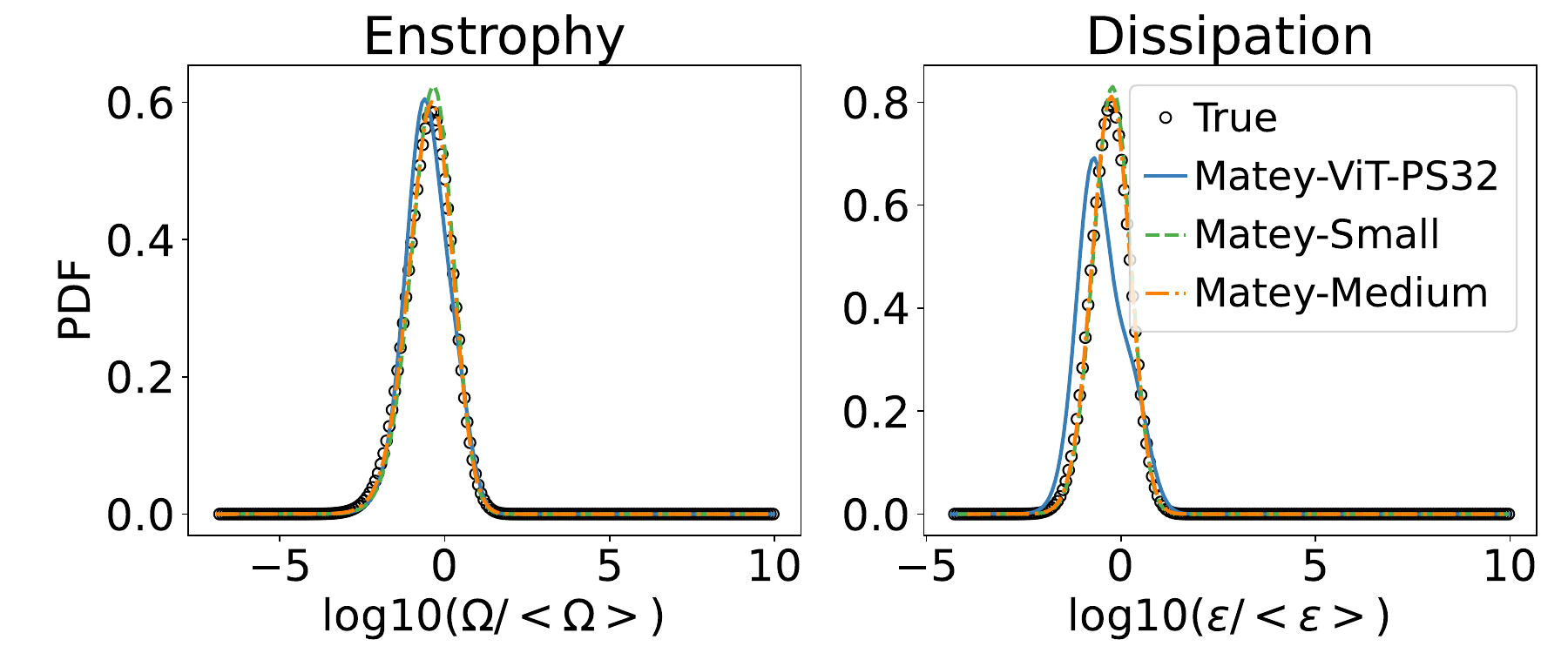}
\includegraphics[width=0.8\linewidth]{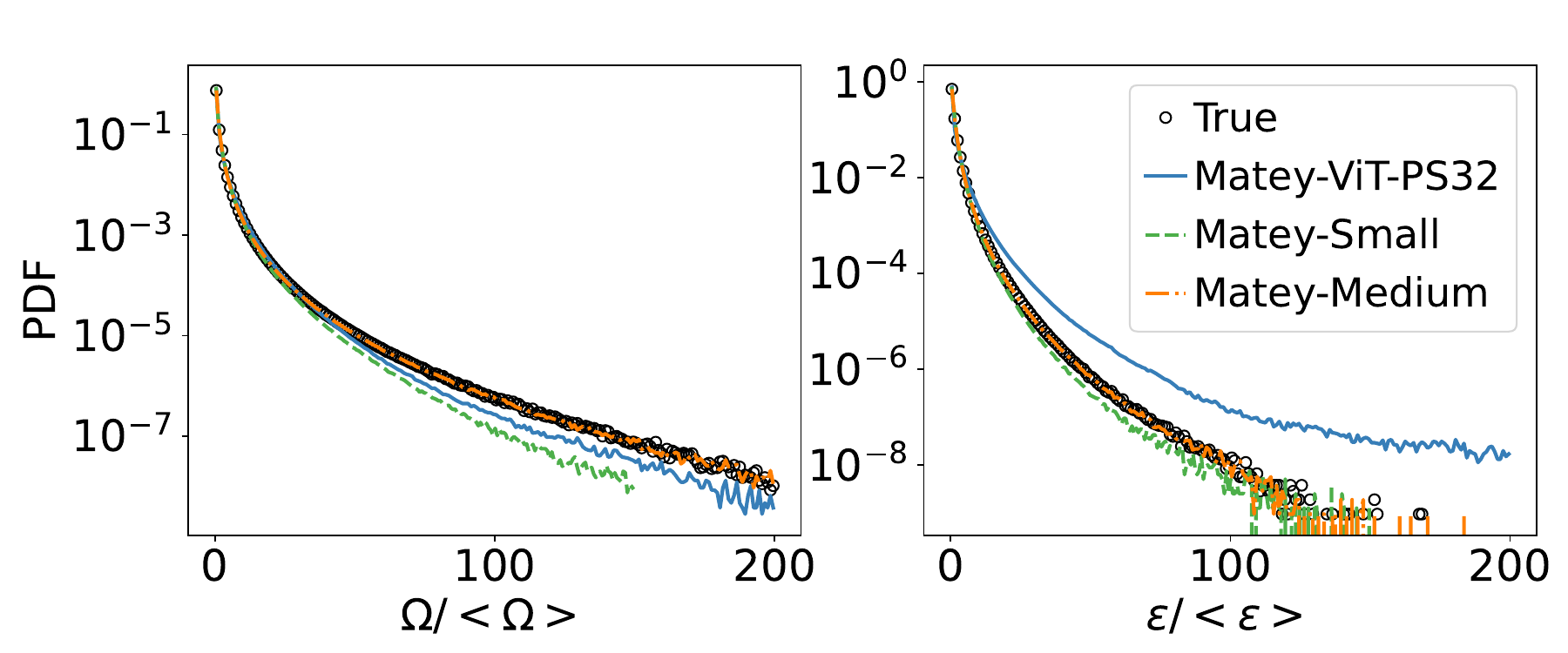}
\caption{Comparison of PDF of enstrophy and dissipation rate, respectively, from DNS and Matey predictions under three configurations.}
\label{fig:omgeps}
\end{figure}

\noindent \textbf{Enstrophy and Dissipation} Fig.~\ref{fig:omgeps} shows the probability density function (PDF) of $\Omega/\langle \Omega\rangle$ and $\varepsilon/\langle\varepsilon\rangle$, respectively, with $\langle \cdot\rangle$ representing spatially averaged values. The top and bottom subplots focus on the peak and tail distributions, respectively. Matey-Medium captures both distributions accurately, including resolving the tail.
This is particularly noteworthy, as capturing the tail is challenging but crucial for understanding turbulence phenomena such as intermittency and high-order moments (e.g., skewness and kurtosis).

\begin{figure}[t]
\centering 
\includegraphics[width=0.8\linewidth]{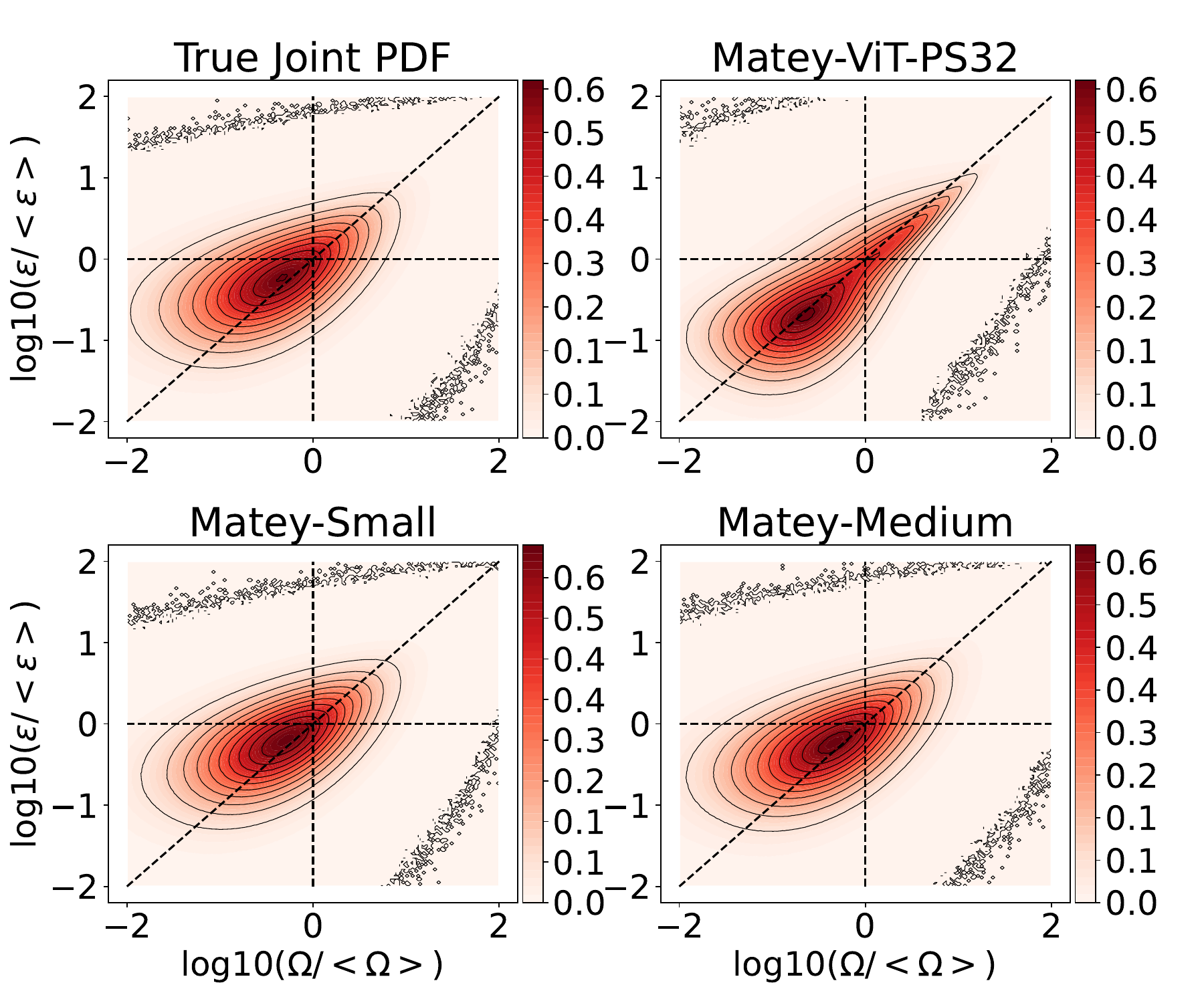}
\caption{Comparison of joint PDF of enstrophy and dissipation rate from DNS and Matey predictions under three configurations.}
\label{fig:jpdfomgeps}
\end{figure}

Fig.~\ref{fig:jpdfomgeps} compares the joint PDF of enstrophy and dissipation from DNS data and the three models. Matey-Medium captures the joint distribution accurately in all four quadrants, while Matey-ViT-PS32 over-predicts the two in the first quadrant at the edge of the high-enstrophy and high-dissipation regime biased toward enstrophy and fails to capture the peak in the third quadrant. These results suggest that our model can make DNS-quality turbulence predictions and hence promise to provide new physical insights together with DNS in future studies. 

\noindent \textbf{Skewness and Kurtosis} Table \ref{tab-skew-kurt} compares the skewness and kurtosis of velocity gradients from DNS data with those predicted by the three models. Negative skewness values represent a forward energy cascade from large to small scales, while high kurtosis values denote heavy-tailed distributions, indicating strong intermittency and extreme events. Both Matey-Small and Matey-Medium reproduce the two quantities satisfactorily, while Matey-ViT-PS32 shows significant deviations. These results further confirm that our models accurately capture small-scale dynamics via the high-order moments of velocity gradients. 

\begin{table}[ht]
\centering
\caption{Comparison of skewness and kurtosis from DNS and Matey predictions under three configurations.}\label{tab-skew-kurt}
\resizebox{0.48\textwidth}{!}{%
\begin{tabular}{c r r r r r r r r}
\toprule
 & \multicolumn{4}{c}{\textbf{Skewness}} & \multicolumn{4}{c}{\textbf{Kurtosis}} \\
\cmidrule(lr){2-5} \cmidrule(lr){6-9}
Component & True & Matey-ViT-PS32 & Matey-Small & Matey-Medium & True & Matey-ViT-PS32 & Matey-Small & Matey-Medium \\
\midrule
$\partial V_x/\partial x$ & -0.525 & -0.061 & -0.530 & -0.487 &  6.985 & 15.279 &  6.777 &  6.993 \\
$\partial V_x/\partial y$ & -0.004 & -0.130 & -0.015 & -0.007 & 10.024 & 18.585 &  8.943 &  9.906 \\
$\partial V_x/\partial z$ &  0.007 &  0.026 &  0.023 &  0.005 & 10.113 & 28.864 &  9.022 & 10.018 \\
$\partial V_y/\partial x$ & -0.017 & -0.103 & -0.027 & -0.013 &  9.792 & 20.599 &  8.762 &  9.625 \\
$\partial V_y/\partial y$ & -0.490 &  0.622 & -0.477 & -0.459 &  6.876 & 34.291 &  6.673 &  6.890 \\
$\partial V_y/\partial z$ &  0.002 &  0.425 &  0.009 &  0.005 &  9.949 & 20.189 &  8.935 &  9.781 \\
$\partial V_z/\partial x$ & -0.041 &  0.355 & -0.025 & -0.038 & 10.073 & 28.777 &  8.855 &  9.984 \\
$\partial V_z/\partial y$ & -0.045 &  0.386 & -0.025 & -0.041 &  9.938 & 18.354 &  8.681 &  9.845 \\
$\partial V_z/\partial z$ & -0.511 & -0.421 & -0.513 & -0.480 &  6.953 & 14.660 &  6.701 &  6.952 \\
\bottomrule
\end{tabular}%
}

\end{table}

\subsection{Stratified Taylor--Green cases}

\begin{figure}[t]
\centering 
\includegraphics[width=0.65\linewidth]{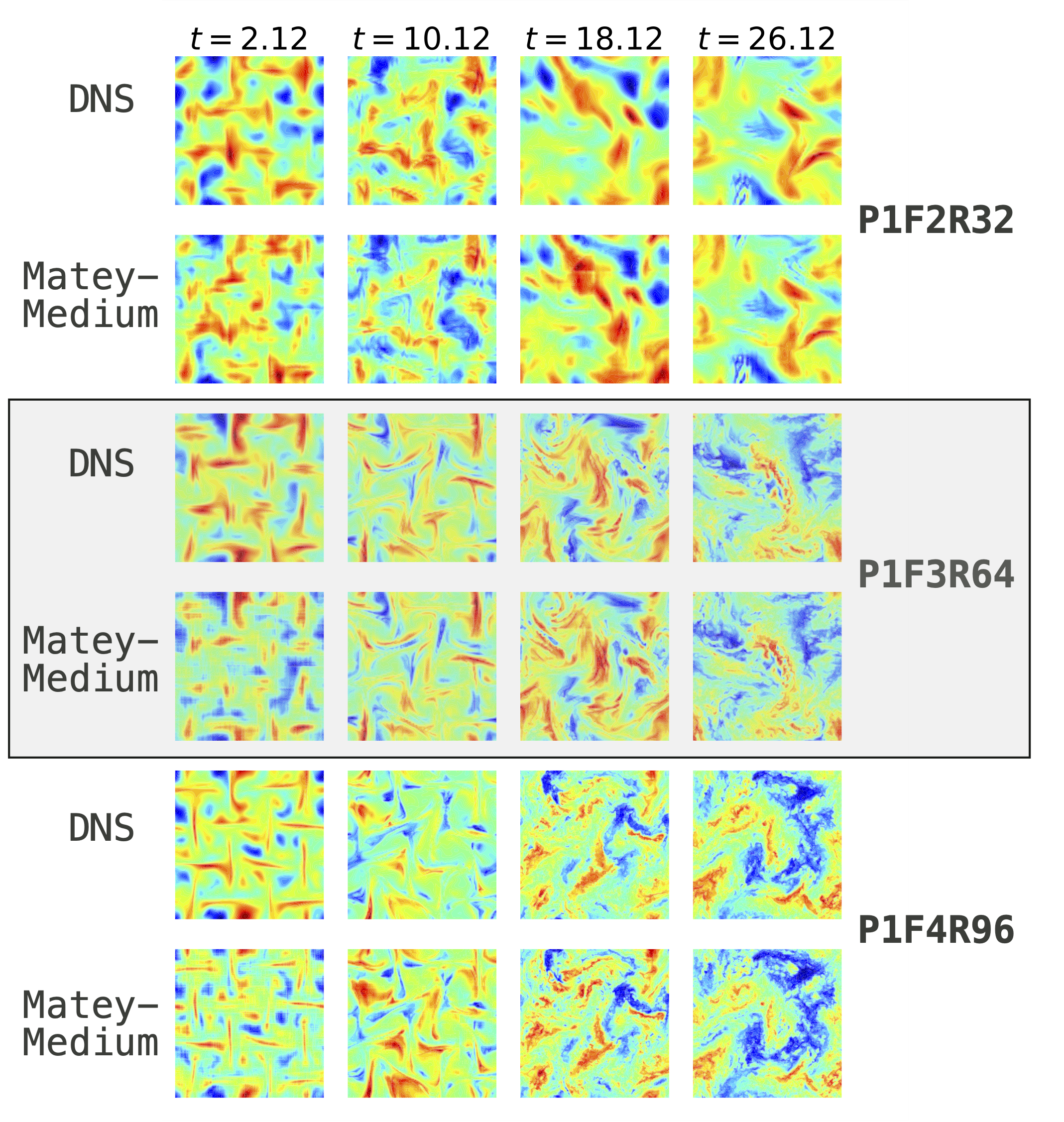}
\caption{Contours of evolving density fluctuations at four selected times for the three test cases from DNS and Matey-Medium predictions.}
\label{fig:tg-rhocontour}
\end{figure}

In this experiment, we evaluate the model's capability in capturing the laminar-to-turbulence transition in a transferable setting beyond the training regimes. The dataset consists of nine cases at varying $\mathit{Fr}$ and $\mathit{Re}$ values with $\mathit{Pr}=1$, as in Table~\ref{tab:tg-param}. Each case consists of 750 snapshots, spanning uniformly a physical time of $[0,30]$. Each snapshot consists of four variables $\textbf{Y}=[\rho^\prime, V_x, V_y, V_z]$, where $\rho^\prime$ is the density fluctuation and the other three are the velocity components. We train the Matey-Medium model to predict the next 10 snapshots, given previous $M=4$ snapshots as input. 
We select the three cases in the diagonal of Table~\ref{tab:tg-param} for testing and the remaining cases for training. Fig.~\ref{fig:tg-rhocontour} compares the predicted density fluctuations in the three test cases with DNS. Matey-Medium captures evolving density well in all three cases.

\begin{table}[ht]
    \centering
    \caption{The grid size [\(N_x, N_y, N_z\)] for nine Taylor--Green cases at varying $\mathit{Fr}$ and $\mathit{Re}$ with $\mathit{Pr}=1$.}
    \label{tab:tg-param}
    \resizebox{0.45\textwidth}{!}{
    \begin{tabular}{lccc}
        \toprule
        & $Fr=2$ & $Fr=3$ & $Fr=4$ \\
        \midrule
        $Re=3200$ & [512, 512, 256] & [512, 512, 256] & [512, 512, 256] \\
        $Re=6400$ & [768, 768, 384] & [768, 768, 384] & [768, 768, 384] \\
        $Re=9600$ & [1024, 1024, 512] & [1024, 1024, 512] & [1024, 1024, 512] \\
        \bottomrule
    \end{tabular}
    }
\end{table}

\begin{figure}[t]
\centering 
\includegraphics[width=1.0\linewidth]{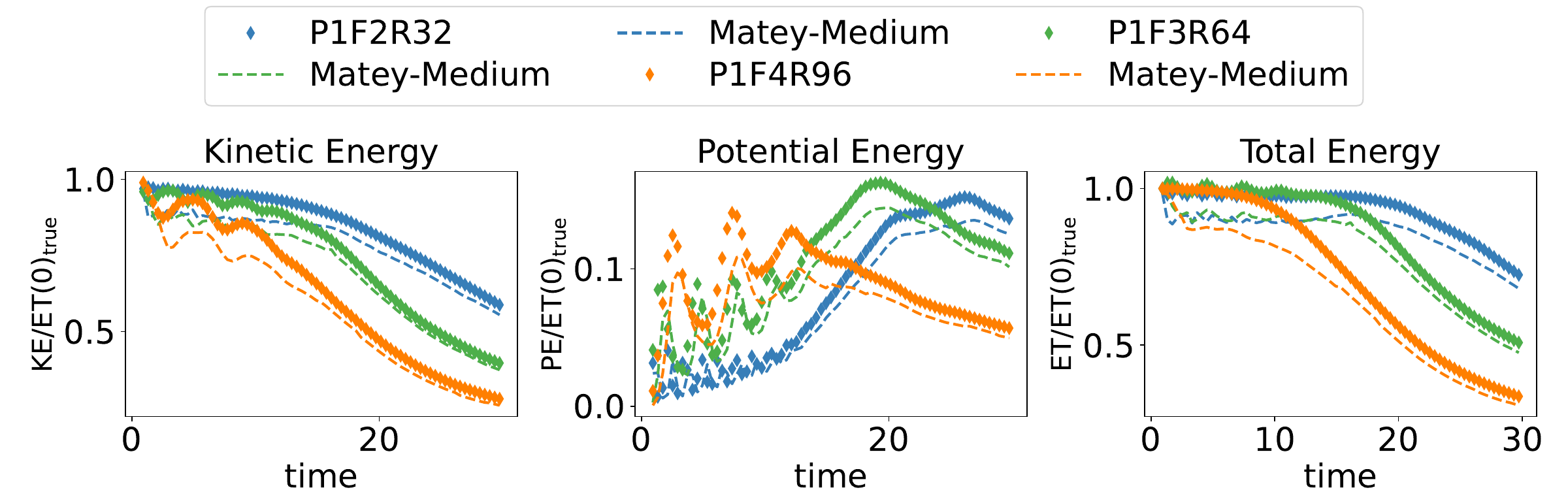}
\caption{Time history of kinetic energy, potential energy, and total energy in three Taylor Green test cases showing laminar to turbulence transition.}
\label{fig:tg-energies}
\end{figure}

Fig.~\ref{fig:tg-energies} shows the time history of spatially averaged kinetic energy, potential energy, and total energy in the three evolving test cases. The data is normalized with the DNS total energy at $t=0$. 
Clearly, Matey-Medium captures the general trend from initial vortex evolution to laminar-turbulence transition and turbulence decay ($\text{time}=[10, 30]$). 
We note that the total energy is under-predicted by the model at a few initial time steps, after which the rate of change of the energy and transition points are well predicted. 
This initial behavior arises because the model does not accurately predict the initial transfer between kinetic and potential energies, which in stratified flow is governed by buoyancy flux ($\propto  {V_z\rho^\prime}$),
 indicating that further investigation is required to capture cross-correlation between the variables. Nonetheless, Matey, enabling accurate generalized prediction of the turbulence transition and the rate of decay of energies, can play a pivotal role in applications such as oceanography and aerodynamics. Such a generalized capability of a model (classical turbulence-theoretic or modern DL-based) is a first to the best of our knowledge.

\section{Implications}

While scientific machine learning has shown remarkable advancement in recent years, addressing more difficult questions---not only in turbulence, but also in areas like omics---will require training on data that are as high-fidelity and high-resolution as possible. Our work suggests a dual-track approach to tackle the  computational challenges in developing DL models on extremely long context data: first, to use known scales and relationships in the data to design hierarchical representations and, second, to develop highly efficient parallel strategies to distribute the computation at each scale across large HPC resources. 

Our RingX approach focuses on the parallelization of the attention kernel, rather than the high-level attention layer (in the case of the sequence parallel in Ulysses \cite{ulysses} and Megatron \cite{megatron}), and hence is compatible with any attention layer design such as multi-head attention (MHA), grouped-query attention (GQA), or multi-head latent attention (MLA). Additionally, RingX is a standalone attention module that can be used as a drop-in replacement for the \texttt{scaled\_dot\_produ\-ct\_attention} in PyTorch. Therefore, it can be easily adopted by Transformer-based applications with minimum effort. Furthermore, RingX eliminates reliance on latency‑sensitive network operations and is positioned to capitalize on future interconnect improvements. 

Our Turbulence Transformer is inspired by the multiscale feature of turbulence, and it reduces the sequence length from the prohibitive billions to a couple of manageable millions, enabling an efficient way to train a model that resolves the smallest features. The method is broadly transferrable to other multiscale spatiotemporal physical systems or, mutatis mutandis, even to the omics data hierarchy in biology. 

Utilizing the two techniques above, Matey achieves near-DNS quality predictions by resolving the finest scales present in the high-resolution turbulence data. 
This breakthrough overcomes the computational bottleneck of training DL models on such data and opens the door for high-fidelity turbulence FMs.
It has three significant scientific implications: 
\begin{enumerate}
    \item Enhanced physical insights: A model that resolves the small-scale eddies and provides high-fidelity predictive capabilities can be used to query the relationships and correlations needed to gain quantitative insights, which would otherwise be only possible through DNS datasets or experimental measurements; 
    \item Continuous and in-situ learning: Because the efficiency of the training strategy and the expressiveness of the model architecture enable ingestion of large-volume DNS data at their highest spatial resolution, continuous learning of a turbulence FM could be performed.  Data can be streamed from DNS to concurrent training  on supercomputers at volumes and temporal resolutions that will far exceed the data capacities of offline training; and 
    \item Broad application impact: A base model for fundamental turbulence---core flow physics shared by diverse application areas like aerodynamics, fusion, combustion, and fire---can be used to derive fine-tuned models for application-specific tasks. Such an approach for developing fine-tuned models that include additional physics at a subsequent stage will allow much more efficient AI/ML solutions than training from scratch.
\end{enumerate}

\section*{Acknowledgment}

This research is sponsored by the Artificial Intelligence Initiative as part of the Laboratory Directed Research and Development (LDRD) Program of Oak Ridge National Laboratory, managed by UT-Battelle, LLC, for the US Department of Energy under contract DE-AC05-00OR22725.
This research was sponsored by and used resources of the Oak Ridge Leadership Computing Facility (OLCF), which is a DOE Office of Science User Facility at the Oak Ridge National Laboratory supported by the US Department of Energy under Contract No. DE-AC05-00OR22725.


\bibliography{refs}
\bibliographystyle{IEEEtran}

\end{document}